\documentclass[12pt]{article}
\usepackage{epsf}
\usepackage{amsmath}
\usepackage{graphicx}
\usepackage{color}
\usepackage{psfrag}

\usepackage{ifpdf}

\newcommand{\bmat}{\left(\begin{array}}
\newcommand{\emat}{\end{array}\right)}

\def\yzero{\smash{\hbox{$y\kern-4pt\raise1pt\hbox{${}^\circ$}$}}}

\def\a{\alpha}
\def\b{\beta}

\def\beq{\begin{equation}}
\def\eeq{\end{equation}}
\def\beqa{\begin{eqnarray}}
\def\eeqa{\end{eqnarray}}

\def\-{\hphantom{-}}
\def\ov{\overline}
\def\s2{\frac{1}{\sqrt2}}

\def\beq{\begin{equation}}
\def\eeq{\end{equation}}
\def\beqa{\begin{eqnarray}}
\def\eeqa{\end{eqnarray}}

\def\IF{\relax{\rm I\kern-.18em F}}
\def\II{\relax{\rm I\kern-.18em I}}

\def\Dsl{\,\raise.15ex\hbox{/}\mkern-13.5mu D} 

\def\IZ{\bf Z}
\def\IT{\bf T}
\def\NN{{\cal N}}
\def\CO{{\cal O}}

\newcommand{\drawsquare}[2]{\hbox{%
\rule{#2pt}{#1pt}\hskip-#2pt
\rule{#1pt}{#2pt}\hskip-#1pt
\rule[#1pt]{#1pt}{#2pt}}\rule[#1pt]{#2pt}{#2pt}\hskip-#2pt
\rule{#2pt}{#1pt}}

\newcommand{\fund}{\raisebox{-.5pt}{\drawsquare{6.5}{0.4}}}

\catcode`\@=11   
\newdimen\@rotdimen
\newbox\@rotbox  

\def\@vspec#1{\special{ps:#1}}
\def\@rotstart#1{\@vspec{gsave currentpoint currentpoint translate
   #1 neg exch neg exch translate}}
\def\@rotfinish{\@vspec{currentpoint grestore moveto}}
\def\@rotr#1{\@rotdimen=\ht#1\advance\@rotdimen by\dp#1%
   \hbox to\@rotdimen{\hskip\ht#1\vbox to\wd#1{\@rotstart{90 rotate}%
   \box#1\vss}\hss}\@rotfinish}
\def\@rotl#1{\@rotdimen=\ht#1\advance\@rotdimen by\dp#1%
   \hbox to\@rotdimen{\vbox to\wd#1{\vskip\wd#1\@rotstart{270 rotate}%
   \box#1\vss}\hss}\@rotfinish}%
\def\@rotu#1{\@rotdimen=\ht#1\advance\@rotdimen by\dp#1%
   \hbox to\wd#1{\hskip\wd#1\vbox to\@rotdimen{\vskip\@rotdimen
   \@rotstart{-1 dup scale}\box#1\vss}\hss}\@rotfinish}%
\def\@rotf#1{\hbox to\wd#1{\hskip\wd#1\@rotstart{-1 1 scale}%
   \box#1\hss}\@rotfinish}%
\def\rotate{\@ifnextchar[{\@rotate}{\@rotate[l]}}
\def\@rotate[#1]#2{\setbox\@rotbox=\hbox{#2}\@nameuse{@rot#1}\@rotbox}

\catcode`\@=12

\topmargin
-1.5cm
\textwidth
15.5cm
\textheight
23.5cm
\oddsidemargin
0.7cm
\evensidemargin
1.2cm

\begin{document}

\makeatletter
\@addtoreset{equation}{section}
\makeatother
\renewcommand{\theequation}{\thesection.\arabic{equation}}
\pagestyle{empty}
\rightline{ IFT-UAM/CSIC-08-50}
\rightline{ CERN-PH-TH/2008-179}
\vspace{0.1cm}
\begin{center}
\LARGE{\bf D-brane instantons and the effective field theory of flux compactifications
  \\[12mm]}
\large{Angel M. Uranga\\[3mm]}
\footnotesize{PH-TH Division, CERN 
CH-1211 Geneva 23, Switzerland\\
 and \\
Instituto de F\'{\i}sica Te\'orica UAM/CSIC,\\[-0.3em]
Universidad Aut\'onoma de Madrid C-XVI, 
Cantoblanco, 28049 Madrid, Spain \\[2mm] }
\small{\bf Abstract} \\[5mm]
\end{center}
\begin{center}
\begin{minipage}[h]{16.0cm}

We provide a description of the effects of fluxes on euclidean D-brane instantons  purely in terms of the 4d effective action. The effect corresponds to the dressing of the effective non-perturbative 4d effective vertex with 4d flux superpotential interactions, generated when the moduli fields made massive by the flux are integrated out. The description in terms of effective field theory allows a unified description of non-perturbative effects in all flux compactifications of a given underlying fluxless model, globally in the moduli space of the latter. It also allows us to describe explicitly the effects on D-brane instantons of fluxes with no microscopic description, like non-geometric fluxes. At the more formal level, the description has interesting connections with the bulk-boundary map of open-closed two-dimensional topological string theory, and with the $\NN=1$ special geometry.

\end{minipage}
\end{center}
\newpage
\setcounter{page}{1}
\pagestyle{plain}
\renewcommand{\thefootnote}{\arabic{footnote}}
\setcounter{footnote}{0}

\vspace*{1cm}

\section{Introduction}

One of the topics of major progress in string theory in recent years is the question of moduli stabilization. The two key mechanisms in this respect are flux compactifications
and non-perturbative effects from wrapped brane instantons (see also \cite{Balasubramanian:2005zx} for scenarios where $\alpha'$ corrections play also an important role). Concerning the first, the simplest flux compactifications, which involve $p$-form field strength fluxes \cite{Dasgupta:1999ss,Giddings:2001yu}, have been generalized to more involved (and less understood) geometric fluxes \cite{Gurrieri:2002wz,Kachru:2002sk} and non-geometric \cite{Shelton:2005cf} flux compactifications. Despite the lack of complete understanding of many aspects of the microscopic description of these general configurations, a uniform description exists in terms of the inclusion of superpotentials (generalizing \cite{Gukov:1999ya}) in the 4d effective field theory action of the fluxless model. The second mechanism, non-perturbative effects from wrapped brane instantons \cite{Becker:1995kb, Witten:1996bn, Harvey:1999as, Witten:1999eg}, has been proposed to contribute non-trivially to the stabilization of those moduli which in certain classes of models are not fixed by flux effects \cite{Kachru:2003aw} \footnote{Other recent activity  in D-brane instantons, initiated in  \cite{Blumenhagen:2006xt,Ibanez:2006da,Florea:2006si}, focuses on the generation of perturbatively forbidden couplings, and its application to potentially phenomenological models.}. Detailed studies of the D-brane instantons contributing to the non-perturbative superpotential in concrete CY compactifications have been carried out, in the absence of fluxes, in  \cite{Denef:2004dm,Denef:2005mm}. These effects are intrinsically four-dimensional, and their computation results in the modifications of the four-dimensional effective action.

The two mechanisms coexist in many of the proposals to achieve complete moduli stabilization, starting with \cite{Kachru:2003aw}. It is therefore crucial to achieve a proper understanding of their interplay, in other words of D-brane instanton effects in flux compactifications. 

Given the difficulties of formulating string theory in the presence of fluxes, a successful viewpoint is to consider the D-brane instantons of the fluxless compactification, and to compute the effects they suffer upon introducing the fluxes. Indeed fluxes can have non-trivial effects on instantons. A drastic example is that some D-brane instantons of the underlying model are not compatible with the flux, due to Freed-Witten anomalies \cite{Freed:1999vc}, and are therefore absent in the flux compactification. This effect has been discussed in \cite{KashaniPoor:2005si}, and we will not focus on it (although it will appear in some of our examples, see section \ref{magnplusnong}). A second kind of effect is that fluxes can lift some of the fermion zero modes of D-brane instantons, therefore modifying the kind of 4d effective interaction they produce. This has been exploited to show that instantons that do not contribute to the superpotential in the absence of fluxes, can contribute to the superpotential in the presence of fluxes.
This effect has been established using different techniques, all relying in a microscopic description of the brane instanton and flux system. For instance the computation of the couplings of worldvolume fermions via the use of the D-brane action coupled to the flux background, \cite{Bergshoeff:2005yp,Tripathy:2005hv,Kallosh:2005gs,Park:2005hj}. A second, recently appeared, technique directly obtains this coupling by computing disk diagrams with two fermion zero mode insertions and one closed string flux vertex operator. Although, when applicable, they lead to the correct result, there are several disadvantages to the microscopic computations of flux effects on instantons. 

\begin{enumerate}
\item Different choices of flux quanta lead to vacua in different corners of moduli space, of the original theory, with presumably very different microscopic flux configuration. In the microscopic approach one needs to study each vacuum individually to establish the 4d effective superpotential appropriate to the corresponding vacuum. 
\item More drastically, the vacuum itself may crucially depend on the non-perturbative effects present in the model. One may have to face the paradoxical situation that the non-perturbative superpotential computed at a point in moduli space forces the model to move away from it (and thus requiring the re-evaluation of the non-perturbative effect in the new candidate vacuum).
\item The microscopic approach to non-perturbative effects of flux compactifications, working pointwise in moduli space, does not make manifest certain intuitions from effective field theory. Consider the effective field theory of the fluxless model, with cutoff the compactification scale, and with a moduli space of vacua. Considering introducing fluxes, with a scale (set up roughly by the flux density) which we take much lower than the compactification scale. The effect of such perturbation must be describable as an effective potential in the moduli space of the original theory, and must therefore admit a global description in moduli space. As we have emphasized above, this is indeed common practice at the classical level (with flux superpotential describing the perturbation in the effective theory), and should be valid for the exact theory including non-perturbative corrections. However, this is not manifest in the microscopic, pointwise in moduli space, approach. 
\item Recent considerations of non-perturbative effects globally in moduli space have led to new insights into their detailed structure, and their continuity and holomorphy properties \cite{GarciaEtxebarria:2007zv,GarciaEtxebarria:2008pi} (see also \cite{Cvetic:2008ws,Gaiotto:2008cd} for related discussions). It is certainly possible that there are similarly interesting questions in the structure of non-perturbative effects in flux compactifications, which cannot be uncovered in the microscopic approach.
\item  Finally, the microscopic approach cannot address the computation of non-perturbative effects in flux compactification without a well understood microscopic description, like non-geometric fluxes, or even certain kinds of geometric fluxes.
\end{enumerate}

In this paper, we propose an alternative approach, which is based on the use of the 4d effective field theory, and does not require a microscopic description of the configuration, so it circumvent the above  difficulties. The basic idea is simple, and follows the line of thought in point 3 above. In the description of the 4d effective action of the original fluxless theory, one must in principle include the contributions from all BPS D-brane instantons, namely the non-perturbative superpotential from instanton with two fermion zero modes, and the multi-fermion F-terms from instantons with additional fermion zero modes (see \cite{Beasley:2004ys,Beasley:2005iu} for such terms in gauge theory and heterotic strings, and \cite{Blumenhagen:2007bn,GarciaEtxebarria:2008pi} in the context of D-brane instantons). Upon introduction of the flux superpotential, some moduli are massive and should be integrated out. This process catalyzes the transformation of some of the original multi-fermion F-terms into new non-perturbative contributions to the superpotential of the flux compactifications (at scales below the moduli mass scale, which is roughly the flux scale). The proposal works globally in moduli space, and does not suffer the problems of a pointwise approach. In addition, it is formulated purely in terms of the 4d effective action, and therefore can address the question of non-perturbative effects in any flux background for which a 4d flux superpotential is known. This includes the familiar $p$-form fluxes, but also geometric and non-geometric fluxes.

The paper is organized as follows. In Section \ref{liftisint} we describe the basic idea in a more quantitative fashion, section \ref{basicidea}, and discuss the procedure in comparison with a more familiar situation in gauge field theory, section \ref{fieldtheory}. 
In Section \ref{exdef} we focus on a concrete example of euclidean D3-branes with fermion zero modes associated to the deformation of the wrapped 4-cycle. In Section \ref{applications} we describe some other examples of IIB D3-brane instantons, focused around the common theme of generating non-perturbative superpotentials from instantons with $U(1)$ Chan-Paton symmetry, by catalysis with 3-form and non-geometric flux superpotentials. In Section \ref{sewing} we use the bulk-boundary map of open-closed topological strings to show the existence of relations between disk and sphere correlators, required for consistency of the microscopic and effective field theory description of flux effects on D-brane instantons. In Section \ref{mixedhodge} we briefly comment on the relevance of ${\NN=1}$ special geometry to D3-brane instantons. Section \ref{conclusion} we make some final remarks. Appendix \ref{bw} contains some properties of multi-fermion F-terms, and appendix \ref{supergraphs} sketches the computation of supergraphs involved in integrating out massive moduli (or dressing up instantons of the original theory by flux interactions).

\section{Lifting fermion zero modes \\ is integrating out massive moduli}
\label{liftisint}

\subsection{The basic idea}
\label{basicidea}

The basic idea in our proposal is that all D-brane instantons induce potentially important terms in the 4d effective action, even in the fluxless compactification. This is clearly recognized for instantons with two fermion zero modes in the fluxless model, since they contribute to the non-perturbative superpotential. But it also holds for D-brane instantons with additional fermion zero modes, which do not contribute to the superpotential, but induce multi-fermion F-terms (denoted higher F-terms in the following), of the kind studied in \cite{Beasley:2004ys,Beasley:2005iu} and reviewed in appendix \ref{bw} (see \cite{Blumenhagen:2007bn,GarciaEtxebarria:2008pi,Matsuo:2008nu,Billo':2008pg} for recent discussions of higher F-terms generated by D-brane instantons). In component terms, these include higher-derivative interactions, and are usually not considered in the effective action of compactifications. However we will argue that they are crucial in providing a consistent description of the introduction of fluxes, since the latter catalyze the transformation of higher F-terms into superpotential terms. 

We first consider a simple model, which will illustrate the main point in a simplified setup, sufficient for our purposes.  Although the discussion is general, for concreteness we phrase it in terms of a type IIB compactification on a CY threefold, with an orientifold projection introducing O3-planes. The compactification contains, in the absence of fluxes, a number of Kahler moduli $T_a$, and of complex structure moduli $U_i$. Consider an euclidean D3-brane instanton wrapped on a 4-cycle $S$ whose volume is controlled by a Kahler modulus $T$. For simplicity we assume that the instanton has a $O(1)$ worldvolume Chan-Paton symmetry. Then, its sector of universal fermion zero modes contains only the translational goldstones $x^\mu$ and the goldstinos $\theta$, and the orientifold projects out the additional fermion zero modes from the $\NN=2$ of the underlying CY compactification). In addition, there are a number of $2n$ fermion zero modes ${\ov \eta}_{i,{\dot \alpha}}$, associated to the deformations of the 4-cycle $S$ invariant under the orientifold projection. Hence ${\ov i}=1,\ldots, n=H_+^{2,0}(S)$, and $\dot\alpha$ labels the two modes, which transform as a chiral spinor of the 4d Lorentz symmetry. There are also bosonic zero modes related to deformations of $S$, but we will ignore them in this discussion. For simplicity we consider $n=1$, since the generalization, briefly mentioned at the end, is straightforward.

In order to compute the 4d amplitude of the D3-brane instanton, we have to integrate over its zero modes, so we need to soak up the fermion zero modes to get a non-zero result. The deformation fermion zero modes can be saturated using the the interaction terms
\beqa
S_{\rm  mix}\,=\, {\ov \eta}_{\dot\alpha}\, {\ov D}^{\dot\alpha}{\ov U}
\eeqa
These interactions of deformation fermion zero modes and complex structure moduli 
will be further discussed in section \ref{sewing}. Note also that they are mirror to those argued for in \cite{Blumenhagen:2007bn}, between deformation fermion zero modes on D6-branes and Kahler moduli in IIA compactifications.

The resulting 4d amplitude for the D3-brane instanton is sketchily of the form
\beqa
S_{4d,0} \, =\, \int \, d^4x \, d^2\theta \, d^{2} {\ov\eta}\, e^{-S_{\rm inst}}\, \simeq\, \int \, d^2\theta\, e^{-T} \, {\ov D}{\ov U}  \, {\ov D}{\ov U}
\eeqa
Our basic statement is that the inclusion of these kind of terms allows a consistent description, purely in terms of the effective action, of the non-perturbative effects in all conceivable flux compactifications of the model. Note that this last property implicitly exploits the fact that the non-perturbative contributions are well-behaved globally in moduli space \cite{GarciaEtxebarria:2007zv,GarciaEtxebarria:2008pi}. 

As a prototypical example, consider introducing NSNS and RR 3-form fluxes in this compactification. Their effect in the effective action is the introduction of the flux superpotential \cite{Gukov:1999ya}
\beqa
W_{\rm flux}\, =\, \int_{X} \, G_3\wedge \Omega
\label{supogvw}
\eeqa
where $G_3=F_3-\tau H_3$, $\tau$ is the complex IIB dilaton, and $\Omega$ is the holomorphic 3-form of the CY threefold $X$. This superpotential can be argued to be exact, even when non-perturbative effects are included, by using the domain wall arguments in the derivation in \cite{Gukov:1999ya}. The only relevant information about this superpotential, for our simplified discussion, is that it generically stabilizes all complex structure moduli. The effective 4d action for the flux compactification thus contains the F-term piece
\beqa
S_{4d,\rm flux} \, =\ \int \, d^4x\, d^2\theta\, e^{-T} \, {\ov D}{\ov U}  \, {\ov D}{\ov U} \,+ \, \int\, d^4x\, d^2\theta \, W_{\rm flux}(U)
\eeqa
Since the superpotential stabilizes $U$ and makes it massive, it can be integrated out to provide an effective action below the scale set by the fluxes. This amounts to computing the Feynman diagrams where the field $U$ runs in internal lines, and does not appear in external lines. The one relevant for us corresponds to an effective interaction from a diagram involving the D3-brane instanton vertex with the two legs of the $U$ field contracted with a superpotential interaction. The diagram can be computed in a manifestly supersymmetric fashion using supergraphs, as we do (in a slightly more general setup) in appendix \ref{supergraphs}.
The resulting 4d interaction, in the effective theory below the flux scale, is
\beqa
S_{4d,{\rm fin}} \, \simeq\, \int \, d^2\theta\, e^{-T} \,  G
\label{finalsupo}
\eeqa
where $G$ is defined by expanding $W_{\rm flux}(U)=GU_2+\ldots$ and is related to the flux density. Hence, the procedure automatically produces the appearance of non-perturbative superpotential from instantons with additional fermion zero modes in the fluxless case.

Some comments are in order:
\begin{itemize}
\item It may seem that we have generated a superpotential from a computation of Feynman diagrams, and that this violates familiar non-renormalization theorems. In fact there is no contradiction, since the final superpotential can be regarded as the original one, expressed in terms of the left-over fields after integrating out the massive ones. 
Related to this, the familiar non-renormalization theorems do not hold in the presence of higher F-terms, which can precisely turn into superpotential terms when suitably dressed by additional interactions.

\item Notice the interpretation of the additional factor of $G$ from the viewpoint of matching of scales. Let us associate a dynamical scale $\Lambda_{\rm n.p.}=e^{-T}$ to the non-perturbative effects of the original theory. The non-perturbative effects of the final effective theory are controlled by a dynamical scale $\Lambda'=e^{-T}G$. This is the correct scale obtained by matching when one integrates out a massive field with mass controlled by $G$.
\end{itemize}
The result (\ref{finalsupo}) matches the expectations from the microscopic approach. Indeed, from the microscopic D3-brane instanton viewpoint, the effect of the flux is to produce a lifting of the fermion zero modes, namely a modification of the D3-brane instantons worldvolume action by a term
\beqa
S_{\rm lift.}\, =\, G {\ov \eta}{\ov \eta}
\eeqa
where we are using the same $G$ as above, for reasons to be discussed later on (see example in section \ref{exdef}, and a general discussion in section \ref{sewing}). Using this interaction to saturate the integration over fermion zero modes, the D3-brane instanton leads to a non-perturbative superpotential in the flux compactification of the above form.

Hence we recover the same non-perturbative superpotential, but exclusively from considerations of the 4d effective action. This is achieved thanks to the inclusion, in the original fluxless 4d effective action, of higher F-terms arising from instantons which do not contribute to the superpotential of the original model. Our work shows that these higher F-terms are nevertheless of crucial importance to allow a systematic discussion of non-perturbative superpotentials in flux compactifications.

\medskip

To illustrate the advantage of the 4d effective theory viewpoint on the interplay of fluxes and instantons, consider the 4d superpotential in the flux compactification. From the microscopic viewpoint, one may be tempted to consider the moduli dependent superpotential to include {\em both} the flux superpotential $W_{\rm flux}$ {\em and} the non-perturbative superpotential $e^{-T}G$. Our derivation above however shows that
this procedure is incorrect, and that these superpotentials belong to two different effective theories: $W_{\rm flux}$ is the superpotential for moduli above the scale set by the flux, so that moduli are still included in the effective action, and the proper non-perturbative effect is the higher F-term $e^{-T} ({\ov {DU}})^2$. And $e^{-T} G$ is the effective superpotential below the flux scale, where the massive moduli have been integrated out. 
Incidentally, note that in a flux compactification where the vacuum expectation value of the superpotential at the minimum $W_0$ is non-zero (hence there is a non-zero cosmological constant in supergravity), the effective superpotential below the flux scale contains the constant piece $W_0$, in addition to the non-perturbative piece $e^{-T}G$. 
Therefore this validates the kind of procedure described in \cite{Kachru:2003aw}.
These issues, which may be confusing from the microscopic viewpoint, become clear in a formulation which allows the deeply rooted intuitions of effective field theory. 

The generalization of our derivation to additional fermion zero modes is straightforward. For an instanton with $2n$ additional fermion zero modes, the effective action of the fluxless compactification contains a non-perturbative higher F-term with $2n$ chiral derivatives of the relevant moduli. Once a flux superpotential is turned on, the effective non-perturbative effects in the flux compactification are obtained by integrating out the massive moduli. Namely by computing Feynman diagrams where the original instanton is dressed by superpotential interactions, which soak up pairs of external legs involving the massive moduli. The supergraphs computation of these diagrams follows the pattern in appendix \ref{supergraphs}. Also, although we have illustrated the idea using D3-brane instantons and deformation zero modes, there are clear generalizations to any other kind of D-brane instanton and of additional fermion zero modes, see later for examples.

\subsection{A simple field theory analog}
\label{fieldtheory}

In this section we consider a simple supersymmetric field theory system, which provides a very precise illustration of the above procedure, but in a more familiar setting. 

Consider four-dimensional $\NN=1$ $SU(N_c)$ SQCD with $N_f=N_c$ flavors. As studied in \cite{Beasley:2004ys}, this theory contains an instanton generating a four-fermion F-term. Namely, there are $2N_c$ neutral fermion zero modes from the gauginos, and $2N_f$ from the flavor quarks. The Yukawa couplings pair up $2N_c-2$ of the "gaugino" fermion zero modes with $2N_f-2$ of the "flavor" fermion zero modes, which are thus all lifted. The left-over "gaugino" fermion zero modes are the goldstinos $\theta^\alpha$ of the two supersymmetries broken by the instanton, while the two "flavor" fermion zero modes ${\ov \eta}$, ${\ov{\tilde \eta}}$ must be saturated against external insertions of the 4d fermion fields. In other words, from the viewpoint of the instanton worldvolume, there is a coupling of the form
\beqa
S_{\rm mix}\, =\, {\ov\eta}\, {\ov D}{\ov Q}\, +\, {\ov{\tilde \eta}}\, 
{\ov D}{\ov {\tilde Q}}\
\eeqa
where actually $Q$, ${\tilde Q}$ are actually combinations of the different flavor fields
in the fundamental/antifundamental. This implies that the instanton leads to an F-term, roughly of the form
\beqa
S_{4d,0}\, =\, \int \, d^4x\, d^2\theta\, e^{-T} \, {\ov D}{\ov Q}\, {\ov D}{\ov {\tilde Q}}
\eeqa
where we have ignored the presence of a non-holomorphic function of the flavor fields.

Consider now perturbing the theory by the introduction of a mass term $m$ for one of the flavors. From the microscopic instanton point of view, this is perceived as the lifting of the two fermion zero modes ${\ov\eta}, \ov{\tilde\eta}$, since they are just modes of the 4d flavor fields. Therefore we have a lifting term on the instanton action
\beqa
S_{\rm lift}\, =\, m {\ov\eta}{\ov{\tilde \eta}}
\eeqa
Saturating the two extra fermion zero modes with this interaction, we see that the instanton amplitude leads to a 4d superpotential term of the form
\beqa
S_{4d,{\rm fin}}\, =\, \int \, d^4x\, d^2\theta\, e^{-T}\, m
\eeqa
where we are ignoring a (holomorphic) function of the flavor fields.
In fact this agrees with the analysis in  \cite{Beasley:2004ys}. There it was argued that the presence of the extra superpotential modifies the supersymmetry algebra, and that the amplitude of the instanton in the deformed theory turns into
\beqa
S_{4d,{\rm BW}}\, =\, \int \, d^4x\, d^2\theta\, e^{-T} \, (\, {\ov D}{\ov Q}\, {\ov D}{\ov {\tilde Q}}\,+\, m\, )
\eeqa
which indeed contains the term of interest.

The above argument is completely analogous to the discussion of lifting of fermion zero modes of D-brane instantons by closed string fluxes. The analog of the flux is the mass parameter, which from the viewpoint of the instanton lifts some of the fermion zero modes, and allows the generation of a non-perturbative superpotential.

Clearly, the field theory example makes it manifest that there is an alternative and more familiar way to interpret these effects, in terms of the 4d effective field theory. The mass deformation makes one of the flavors massive, so that integrating it out we are left with $N_f=N_c-1$ SQCD. The instanton in this theory generates an Affleck-Dine-Seiberg superpotential. In more detail, after the introduction of the mass parameter, the theory has the F-terms
\beqa
S_{4d,{\rm mass}}\, =\,  \int d^4x\, d^2\theta \, e^{-T}\, {\ov D}{\ov Q}\, {\ov D}{\ov {\tilde Q}}\, +\,   \int d^4x\, d^2\theta \, m\, Q\,{\tilde Q}
\eeqa
Integrating out the massive flavor amounts to evaluating the dressed instanton diagrams, with two external flavor legs contracted with a mass term, similar to the diagram \ref{fig:supergraph}. The resulting dressed instanton interaction gives a superpotential term
\beqa
S_{4d,{\rm fin}}\, =\,  \int d^4x\, d^2\theta \, e^{-T}\, m
\eeqa
in the effective $N_f=N_c-1$ SQCD theory. Restoring a (holomorphic) function of the 
flavors, and using matching of scales, this reproduces the expected ADS superpotential. Thus the instanton superpotentials in the deformed effective theories can be obtained by dressing the instantons of the original theory with the deformation interaction. Our procedure in Section \ref{basicidea} is simply the application of this effective field theory logic to D-brane instantons in flux compactifications.

\medskip

The above field theory model can be easily engineered into string theory, e.g. as in \cite{GarciaEtxebarria:2008pi} (see also \cite{Billo':2008pg}). Equivalently, it can be done in the T-dual Hanany-Witten picture \cite{Hanany:1996ie,Elitzur:1997fh} (see \cite{Giveon:1998sr} for a review) with color D4-branes stretched between two relatively rotated NS-branes, and semi-infinite D4-branes yielding the flavors. the analogy with the discussion in the previous section can be made completely precise, by applying a T-duality similar to that in \cite{Cascales:2003ew}, as follows: The NS-branes turn into a conifold-like singularity, the D4-branes turn into D3-branes, and the mass term (corresponding to a twist in the original HW geometry \cite{Witten:1997sc}) turns into a background 3-form flux. We refrain from entering additional details of this realization, and content ourselves with stating the complete analogy of the procedures in the previous section with standard practice in supersymmetric gauge field theory.

\section{Example: deformation fermion zero modes and 3-form fluxes}
\label{exdef}

In this Section we consider more explicit examples of the above mechanism, focusing on D3-brane instantons with deformation fermion zero modes, in the presence of 3-form fluxes.

\subsection{The setup}

Let us consider type IIB theory compactified on a CY threefold $X$. Since we are mostly interested in the ${\cal N}=1$ supersymmetric case, we consider there are orientifold planes, which for definiteness we take to be O3/O7 planes (analogous discussion could be carried out for M5-brane instantons in compactifications of F-theory on CY fourfolds). Still, since we are interested in instantons which may have extra fermion zero modes we do not assume that they are mapped to themselves by the orientifold action, and so the orientifold projection is not very crucial. In any event, the fact that we eventually turn on flux superpotentials makes it natural to use the N=1 language.

Consider an euclidean D3-brane instanton wrapped on a holomorphic 4-cycle $S$. For simplicity we consider its worldvolume gauge bundle to be trivial.  For a $U(1)$ instanton, in addition to the universal set of fermion zero modes, we have sets of deformation bosonic zero modes $\phi$, and their fermionic partners $\eta^\alpha$. The index $\alpha$ denotes a spinor index in spacetime, so we have two fermion zero modes of this kind. They are associated to deformations of the 4-cycle $S$ keeping it holomorphic. Namely to sections of the holomorphic normal bundle of $S$ on $X$. For each such section $\phi^a\in H^0(N_{S})$ we can use the holomorphic $(3,0)$ form $\Omega$ on $X$ to construct $\phi_{bc}=\phi^a \Omega_{abc}$, which is an element of $H^{(2,0)}(S)$. So the deformation zero modes of the D3-brane instantons are classified by the latter group (at least locally in moduli space, see section \ref{mixedhodge} for a discussion about issues more globally in moduli space). We are interested in a $U(1)$ instanton with $2p$ of these fermion zero modes, with $p=h^{2,0}(S)$. For simplicity we assume that the instanton has $O(1)$ Chan-Paton symmetry, so that the universal sector contains only two fermion zero modes, the goldstinos $\theta$.

In the following we particularize this general analysis to simple flat factorized examples, like $\IT^4\times \IT^2$ or  K3$\times \IT^2$, realized as $(\IT^4/\IZ_2)\times \IT^2$ (or additional quotients thereof), with a D3-brane instanton wrapped on the fiber $\IT^4$ or K3, and sitting on top of the orientifold plane. We choose the latter so that the D3-brane 
has Chan-Paton symmetry $O(1)$ and the deformation fermion zero mode survives the orientifold projection.

\subsection{The microscopic viewpoint: Lifting zero modes by fluxes}
\label{microview}

Let us consider the effect of fluxes on these fermion zero modes, and how they are lifted. In order to keep the discussion as clean as possible, let us consider a supersymmetric vacuum of the theory, with a primitive $(2,1)$ flux density $G_3$. The lifting of fermion zero modes of D3-branes by fluxes has been studied in several works, and we simply borrow the relevant results. 
The coupling can be computed using the D-brane action coupled to a general supergravity background \cite{Bergshoeff:2005yp} (see also \cite{Tripathy:2005hv,Kallosh:2005gs,Park:2005hj}, or as we discuss later by computing correlators in the fluxless theory. The relevant piece for us has the structure
\beqa
({\tilde G}_3)_{{\hat m}{\hat n}p}\, \Theta \, \Gamma^{{\hat m}{\hat n}p}\, \Theta
\eeqa
where $\Theta$ encodes all worldvolume D3-brane fermions (expressed as a 10d spinor, or as spinors on the D3-brane transforming also as spinors of the normal bundle in the CY and of the 4d spacetime directions), and hatted/unhatted coordinates correspond to directions parallel/transverse to the D3-brane instanton. Also ${\tilde G}_3$ is related to the continuation of the standard $G_3$ to euclidean signature. More specifically
\beqa
{\tilde G}_3\, =\, i\, (\, F_3\, +\, {\tilde \tau} \, H_3\,) \quad {\rm with} \; {\tilde\tau} \, =\, i \, e^{-\phi}\,+\, a\, \gamma_5
\eeqa
Since we focus on a particular spinor with fixed eigenvalue of the D3-brane worldvolume chirality $\gamma_5$, ${\tilde G}_3$ essentially reduces to the standard $G_3$, modulo a proportionality constant.  

This approach has been often applied in the literature, and we will not repeat it here. Instead for future convenience it will be useful to use a different derivation of the coupling for our case of interest. It is based on the study of the effect of type IIB 3-form fluxes on D-brane instanton fermion zero modes as carried out in \cite{Billo':2008sp}. The approach is to compute disk correlators involving two fermion zero modes and one vertex operator of the closed string 3-form fluxes. We refer to these references for details, and restrict the discussion to selection rules, based on H-momentum in the bosonized formalism, which are sufficient to show the existence of the relevant couplings. 

We consider a 3-form flux which preserves the supersymmetry of the vacuum, and in particular is of the form 
\beqa
G_3\, =\,G\, dz^1\, dz^2\, d{\ov {z^3}}
\label{theflux}
\eeqa
This corresponds to a weight vector $(++-)$ of the internal CFT. 
The deformation fermion zero modes are associated to the $(2,0)$ form $dz^1dz^2$ on the 4-cycle, and thus correspond to weight vectors
\beqa
\frac 12 (++-;--) \quad ,\quad \frac 12 (++-++)
\label{fzms}
\eeqa
where we have also shown the 4d spacetime spinor weights, which are implicit in what follows.
There is a cubic coupling corresponding to the correlator structure
\beqa
\langle \, (++-)_{G_3}\,\cdot\,  \frac 12\, (--+)_{\, \ov \eta}\,\cdot \,  \frac 12\, (--+)_{\, \ov \eta} \, \rangle
\eeqa
This leads to a coupling
\beqa
S_{\rm lift}\, =\, G\, {\ov \eta}\,{\ov \eta}
\eeqa
Thus in the presence of these flux, the D3-brane instanton provides a 4d superpotential
\beqa
S_{4d,\rm fin}\, =\, \int d^4x \, d^2\theta\, G\, e^{-T}
\label{nfluxsupo}
\eeqa

\subsection{Effective field theory viewpoint: integrating out moduli}
\label{eftview}

In the following we show how this result is derived from our effective field theory approach. Consider the fluxless compactification. The D3-brane instanton considered above has one set of additional deformation fermion zero modes. These have couplings to complex structure moduli, as follows
\beqa
S_{\rm mix}\, =\, {\ov \eta}\, (\, {\ov {DU_1}}\, +\, {\ov {DU_2}})
\label{defcom}
\eeqa
That these couplings are allowed, can be seen by using the H-momentum structure of the 4d modulinos
\beqa
& \psi_{U_1}\, \leftrightarrow & 
\frac 12\, (+-+)_L\, \otimes\, (-00)_R\, + \, {}_{(L\leftrightarrow R)} \nonumber \\
& \psi_{U_1}\, \leftrightarrow & 
\frac 12\, (-++)_L\, \otimes\, (0-0)_R\, + \, {}_{(L\leftrightarrow R) }
\label{modulinos}
\eeqa
where there is a left-right symmetrization to form combinations invariant under the orientifold action.
So there are allowed correlator structures
\beqa
& \langle\, \frac 12\,  (--+)_{\,\ov \eta} \, \cdot\, \left( \,  \frac 12\, (-+-)_L\, \otimes\, (+00)_R\, + \, {}_{(L\leftrightarrow R)} \, \right)_{\ov{\psi}_{U_1}} \, \rangle \nonumber \\
& \langle\, \frac 12\,  (--+)_{\,\ov \eta} \, \cdot\, \left( \,  \frac 12\, (+--)_L\, \otimes\, (0+0)_R\, + \, {}_{(L\leftrightarrow R)} \, \right)_{\ov{\psi}_{U_2}} \, \rangle 
\eeqa
Note that the boundary conditions of the worldsheet fields on the disk relate the left and right pieces of the 4d modulino vertex operator, allowing non-vanishing correlators.
It is worthwhile to point out that the correlators are also non-vanishing in the unorientifolded theory, namely for instantons not mapped to themselves under the orientifold action.

Besides the symmetry between the first two complex planes, there is a heuristic argument to show the coupling of the deformation fermion zero modes to the combination of complex structure moduli as in (\ref{defcom}). Consider momentarily the theory in the absence of orientifold plane. Then the closed string background has a hyperkahler structure, and a triplet of kahler forms $J_i$. The wrapped D3-brane selects a particular complex structure, with respect to which the forms organize as a Kahler and complex structure 2-forms on K3 or $T^4$ as $J=J_3$, $\Omega=J_1+iJ_2$. In the latter case we can write
\beqa
J\,  =\, dz^1\, d{\ov{z^1}}\, +\, dz^2\, d{\ov {z^2}} \quad , \quad \Omega\, =\, dz^1\, dz^2
\eeqa
In the complete threefold, they can be used to construct the $(2,1)$ forms $J\, dz^3$, associated to $U_1+U_2$, and $\Omega\, d{\ov {z^3}}$, associated to $U_3$. The structure of the coupling of the worldvolume fermions to $U_1+U_2$ in inherited from this underlying property.

Using the couplings (\ref{defcom}), the D3-brane instanton generates a higher F-term in the fluxless compactification, with the structure
\beqa
S_{4d,0}\, =\, \int\,d^4x\,  d^2\theta \, e^{-T}\, \left(\, {\ov D}{\ov U}_1\, +\, {\ov D}{\ov U}_2\, \right)^2 
\eeqa

Upon introduction of fluxes, the theory is deformed by the flux superpotential. We are interested in the term coupling the $(2,1)$ flux (\ref{theflux}) to the 4d modulinos in the chiral multiplets $U_1$, $U_2$. These can be computed from a correlator in the sphere, and can be argued to be allowed by the selection rules. Indeed using (\ref{modulinos}), we find allowed correlators
\beqa
\langle\, (++-)_{G_3}\cdot
  \big(\, \frac 12\, (+-+)_L\, \otimes (-00)_R\, +  {}_{(L\leftrightarrow R)} \, \big)_{\psi_{U_1}} \cdot \big(\,  \frac 12\, (-++)_L\, \otimes (0-0)_R + {}_{(L\leftrightarrow R)} \, 
  \big)_{\psi_{U_2}}\,  \rangle \nonumber
\eeqa
Hence there is a component of the superpotential
\beqa
W_{\rm flux}\, =\, G\, U_1\, U_2
\label{termgvw}
\eeqa
Integrating out the moduli stabilized by the flux, and taking into account their appearance in the higher F-term in the original effective action, we can derive the non-perturbative superpotential (\ref{nfluxsupo}). Notice the advantage that the effective action for the fluxless compactfication can be used for any flux which can conceivably be turned on. This can be used to provide some non-trivial examples of non-perturbative superpotentials in quite involved flux environments, involving for instance non-geometric fluxes, see section \ref{magnplusnong}.

Before that, we would like to mention that the term (\ref{termgvw}) can also be recovered following the more standard form of the 4d flux superpotential, namely (\ref{supogvw}).  We are interested in the quantity
\beqa
\frac{\partial^2 W}{\partial U_1\partial U_2}\, =\, \int_X\, G_3 \wedge \frac{\partial^2\Omega}{\partial U_1\partial U_2}
\label{doublederivative}
\eeqa
evaluated at the minimum of the flux potential. This provides the coefficient of the effective vertex dressing the instanton induced 4d higher F-term. Namely, it plays the role of the coefficient $G$ in the above discussion. The toroidal model is simple enough so that the latter can be computed easily from
\beqa
\Omega\, =\, dz^1\, dz^2\, dz^3\, =\, (dx^1 + U_1 dy^1)\, 
(dx^2 + U_2 dy^2)\,(dx^3 + U_3 dy^3)\,
\eeqa
where $dx^i,dy^i$ for a basis of integer forms in the $i^{th}$ two-torus. The computation can still be further simplified by using some additional general considerations. Note that in Minkowski vacua $G_3$ must have $(0,3)$ or $(2,1)$ components. Each holomorphic derivative of $\Omega$ flips at most one holomorphic leg to an antiholomorphic one, hence $\partial_{12} \Omega$ has $(3,0)$, $(2,1)$ and $(1,2)$ components. Then the $(0,3)$ component of $G_3$ does not contribute to (\ref{doublederivative}). Finally we 
notice that the $(1,2)$ component of $\partial_{12}\Omega$ is
\beqa
\left.  \frac{\partial \Omega}{\partial U_1\partial U_2}\right|_{(1,2)}\, =\, d{\ov{z^1}}\, d{\ov {z^2}}\, dz^3
\eeqa
Thus the only contribution to (\ref{doublederivative}), and hence to the coefficient $G$ in the above discussion, is the piece of $G_3$ proportional to $dz_1dz_2d{\ov{z^3}}$, as used above.

From the viewpoint of the 4d effective action, one may not be particularly interested in the Hodge structure of the different quantities mentioned in the computation, but rather in the result, expressed in terms of the basic inputs of the compactification, namely the flux quanta. This is straightforward to obtain from the standard form of the 4d superpotential used in the study of moduli stabilization, as we now discuss.
The procedure of computing the complete flux superpotential on tori was described in detail in \cite{Kachru:2002he,Frey:2002hf}. For later convenience, we follow the notation in \cite{Aldazabal:2006up}. We introduce a basis of integer homology cycles, obtained by tensoring the forms $dx^i$, $dy^i$, on each two-torus
\beqa
\alpha_0 & = & dx^1 dx^2 dx^3 \quad ; \quad
\beta_0 = dy^1 dy^2  dy^3 \ , \nonumber \\[0.2cm]
\alpha_1 & = & dx^1 dy^2 dy^3 \quad ; \quad
\beta_1 = dy^1 dx^2 dx^3 \ , \label{abbasis} \\[0.2cm]
\alpha_2 & = & dy^1 dx^2 dy^3 \quad ; \quad
\beta_2 = dx^1 dy^2 dx^3 \ , \nonumber \\[0.2cm]
\alpha_3 & = & dy^1 dy^2 dx^3 \quad ; \quad
\beta_3 = dx^1 dx^2 dy^3 \ , \nonumber
\eeqa
In this basis, the fluxes have an expansion
\beqa
F_3 & = & -m \a_0 -e_0\b_0 + \sum_{i=1}^3 (e_i\a_i - q_i \b_i) \nonumber \\
H_3  & = &  h_0\b_0 - \sum_{i=1}^3 a_i\a_i + \bar h_0\a_0 - \sum_{i=1}^3 \bar a_i\b_i   
\label{fluxesacfi} 
\eeqa
Introducing the complex coordinates $z^i=x^i+U^iy^i$, and using $\Omega\, =\, dz^1\, dz^2\, dz^3$,we have
\beqa
W_{\rm flux} & = & \int G_3\wedge \Omega\, =\, e_0 + i\sum_{i=1}^3 e_i U_i
- q_1 U_2 U_3 - q_2 U_1 U_3 - q_3 U_1 U_2 +i m U_1 U_2 U_3
\nonumber \\[0.2cm]
& + & S \big[ ih_0 - \sum_{i=1}^3 a_i U_i
+i \bar a_1 U_2 U_3 +i \bar a_2 U_1 U_3 +i \bar a_3 U_1 U_2 - \bar h_0 U_1 U_2 U_3 \big]
\label{supoexplicit}
\eeqa
where we are using $S=-i\tau$. This superpotential generically makes massive all the complex structure moduli. Using this expression for the superpotential, we have
\beqa
\frac{\partial^2 W}{\partial U_1\partial U_2}\, =\, -q_3 \, +\, im\, U_3\, +i \bar a_3 \, S\, -\bar h_0\, S\, U_3
\label{gfrom4d}
\eeqa
Taking the values of $S, U_3$ at the minimum of the flux potential, we obtain the coefficient of the effective vertex dressing the instanton induced 4d higher F-term. Namely the quantity (\ref{doublederivative}), directly in terms of the flux quanta (and the vevs of $S,U_3$). Thus upon integrating out the fields $U_1,U_2$ as described above, we obtain the 4d non-perturbative superpotential
\beqa
S_{4d,\rm fin}\, =\, \int\, d^4x\, d^2\theta\,  (\, -q_3 \, +\, im\, U_3\, +i \bar a_3 \, S\, -\bar h_0\, S\, U_3 \,)\, e^{-T}
\eeqa
where in general the coefficient must be evaluated at the minimum of the flux potential, if the moduli $S,U_3$ are massive with mass given by the flux scale (since the above superpotential is valid in the effective theory below this scale).

\section{Applications}
\label{applications}

We hope to have convinced the reader about the virtues of the effective field theory approach, and its full generality beyond the concrete example considered above. In order to illustrate better the power of this technique, in this section we consider some further examples, which include fluxes with no 10d description (and for which the microscopic approach is thus beyond present techniques).

Although the examples in this section can be taken on their own, the common theme is the use of our approach to obtain non-perturbative superpotentials from generic D3-brane instantons, not mapped to themselves under the orientifold action. These typically have additional fermion zero modes which are notoriously difficult to lift, and whose lifting is related to supersymmetry breaking in spacetime \cite{GarciaEtxebarria:2008pi}.

\subsection{Revisiting a no-go result: $U(1)$ instantons and 3-form fluxes}
\label{nogo}

The effective field theory approach allows to derive in very simple terms some known results in the literature. For instance, consider a general CY compactification without fluxes, and an euclidean D3-brane wrapped on a rigid 4-cycle, not mapped to itself under the orientifold action, and with no world-volume flux turned on. This instanton has two additional fermion zero modes ${\ov\tau}$, which are goldstinos of the accidental $\NN=2$ supersymmetry in the underlying CY compactification, and therefore does not contribute to the superpotential. Naively, one might think that it is possible to lift these fermion zero modes with suitable 3-form fluxes. However, a microscopic analysis using the D3-brane action coupled to fluxes shows this not to be possible \cite{Blumenhagen:2007bn}.

This has a very simple explanation from the effective field theory point of view.
The instanton is an example of D-brane instanton whose BPS phase can misalign, with respect to the $\NN=1$ susy of the compactification, and which were studied in \cite{GarciaEtxebarria:2008pi}. For such instantons, the two additional fermion zero modes ${\ov\tau}$ have couplings to the Kahler modulus $\Sigma$ which controls the BPS misalignment, and thus lead to an effective 4d higher F-term contribution of the form
\beqa
S_{4d,0}\, =\, \int\, d^4x\,  d^2\theta\, e^{-T}\, {\ov{D\Sigma}}\, {\ov{D\Sigma}}
\eeqa
Turning on 3-form fluxes leads to the deformation of the theory by the superpotential (\ref{supogvw}), which depends only on the dilaton and complex structure moduli. Therefore, it is not suitable to dress the D3-brane instanton amplitude and turn it into a superpotential. The above higher F-terms remains intact in this class of flux compactifications.

It is very satisfactory to check that the effective field theory approach provides new intuitions concerning results which do not have an obvious physical interpretation in the microscopic picture. 

\subsection{Magnetized $U(1)$ instantons}
\label{magnetized}

One may suspect that the no-go result is just an accident of the particular kind of flux compactification considered. In fact, it has been recently shown in \cite{Billo':2008sp,Billo':2008pg} that 3-form fluxes can lift similar fermion zero modes in D$(-1)$-brane instantons. Performing suitable dualities, we would conclude that it should be possible to lift these fermion zero modes for D3-branes by using non-geometric fluxes. 
In previous approaches it was not possible to verify this directly. However, we can use the effective field theory techniques to show this in concrete examples. In order to do so, it will be convenient to have concrete examples of $U(1)$ instantons.

Let us consider the closed string background of section \ref{exdef}, but focus on a different D3-brane instanton. We again focus on instantons wrapped on the K3 (or $\IT^4/\IZ_2$) fiber, but consider turning on a worldvolume magnetic flux. For simplicity, we keep the factorized structure of the tori, so
\beqa
F\, =\, m_1 \, dx^1\,dy^1\,+\, m_2 dx^2\, dy^2
\eeqa
In order to be BPS, the worldvolume flux must be primitive. Hence the instanton is BPS on the locus $m_1T_1+m_2T_2=0$, and become non-BPS away from it. 
This is a simple example of D-brane instanton whose BPS phase can misalign, and which were studied in \cite{GarciaEtxebarria:2008pi}. For concreteness, we focus our analysis on the BPS locus in Kahler moduli space.

Such instantons generate higher F-terms involving derivatives of Kahler moduli. Indeed, due to the worldvolume magnetic flux, the D3-brane instanton is not mapped to itself under the orientifold action, and has worldvolume gauge group $U(1)$ (which thus supports the flux). The universal sector of zero modes contains two additional fermion zero modes ${\ov \tau}$ (which become the extra goldstinos ${\ov \theta}$ away from the BPS locus). These fermion zero modes have couplings with the 4d moduli 
controlling the BPS phase of the instanton \cite{GarciaEtxebarria:2008pi}, namely 
\beqa
S_{\rm mix}\, =\, {\ov \eta}\, {\ov D}\, (\, m_1{\ov T_1}\,+\, m_2\,{\ov T_2}\,)
\eeqa
Note that in this formula we are restricting to the coupling to the factorized Kahler moduli. In general, there are couplings to additional Kahler moduli (off diagonal in $\IT^4/\IZ_2\times \IT^2$, or twisted in $\IT^6/(\IZ_2\times \IZ_2)$), which are not shown. These couplings would be relevant in the case with no worldvolume magnetic fluxes $m_1=m_2=0$. In other words, we have used magnetization of the D3-brane instanton as a trick to make the couplings of ${\ov\tau}$ to Kahler moduli without going out of the factorized ansatz.

In addition to ${\ov\tau}$, the instanton has the deformation zero modes ${\ov \eta}$ discussed above. In total, the non-perturbative contribution of the instanton has the structure
\beqa
S_{4d,0}\,=\,\int \,d^4x\,  d^2\theta\, e^{-T}\, [\, {\ov D}(\, m_1{\ov T_1}\,+\, m_2\,{\ov T_2}\,)\,]^2\, [\, {\ov D}(\,c_1{\ov U_1}\,+\,c_2{\ov U_2}\,)\,]^2\,
\label{magneticfterm}
\eeqa
where $T$ is a combination of the different Kahler moduli, dependent on $m_1,m_2$, and $c_1,c_2$ are coefficients depending also on the magnetization quanta. These details will not be very relevant to our purposes, and we skip them.

As we have mentioned in different opportunities, it is now straightforward to discuss the possible non-perturbative effects derived from this instanton in different flux compactifications, by simply introducing the relevant superpotential in the effective action and integrating out the massive moduli.

\subsection{Magnetized D3-brane instantons and 3-form fluxes}
\label{magnplus3form}

Just for reference, let us describe the introduction of 3-form fluxes in the compactification and their effect on the above D3-brane instantons. Introducing the superpotential (\ref{supogvw}), in the form (\ref{supoexplicit}), we see that it is possible to use superpotential interactions to contract the external legs of complex structure fields in the instanton amplitude (\ref{magneticfterm}). However, it is not possible to contract the external legs of Kahler moduli. So the instanton non-perturbative effects is a higher F-term of the form
\beqa
S_{4d,{\rm fin}}\, =\, \int \,d^4x\,  d^2\theta\, G\, e^{-T}\, [\, {\ov D}(\, m_1{\ov T_1}\,+\, m_2\,{\ov T_2}\,)\,]^2\,
\eeqa
where $G$ is the coefficient of the $dz^1dz^2d{\ov z^3}$ term in $G_3$, as above.

In microscopic terms, the 3-form fluxes have lifted the fermion zero modes ${\ov\eta}$, but not the ${\ov\tau}$. This system combines features of the instantons considered in section \ref{exdef} and \ref{nogo}. Note that the possible lifting of the ${\ov\tau}$ zero modes of magnetized D3-branes by 3-form fluxes in \cite{Blumenhagen:2007bn} requires the introduction of off-diagonal magnetic fluxes, hence lies outside the class we are considering \footnote{In fact, the analysis of this effect in \cite{GarciaEtxebarria:2008pi} suggests a link between the lifting of the ${\ov \tau}$ fermion zero modes and the breaking of supersymmetry by non-primitivity of $G_3$. This suggests that this lifting may not be achievable in CY compactifications other than $\IT^6$ or K3$\times \IT^2$, and thus may not be describable in an effective action including only the superpotential (\ref{supogvw}).}.

\subsection{Magnetized D3-brane instantons and  non-geometric fluxes}
\label{magnplusnong}

The effective field theory approach allows to consider a possible mechanism to lift the $\ov\tau$ fermion zero modes, by turning on additional non-geometric fluxes. The fact that such fluxes do not have a 10d geometric description prevents the use of a microscopic picture \footnote{For particular choices of flux, it may be possible to dualize the model to a geometric one. Also it may be possible that certain effects are local enough to allow for a local 10d geometric description. See e.g. \cite{Marchesano:2007vw} for studies in this direction.}. However, it poses no obstacle to a treatment in the effective field theory description.

The superpotential for 3-form and non-geometric fluxes has been considered in detail e.g. in \cite{Shelton:2005cf}. For our purposes, the expression for factorized toroidal geometries in type IIB compactifications with O3-planes, we may use the expression in \cite{Aldazabal:2006up}.
\beqa
W_{\rm n.g.flux} & = & e_0 + i\sum_{i=1}^3 e_i U_i
- q_1 U_2 U_3 - q_2 U_1 U_3 - q_3 U_1 U_2 +i m U_1 U_2 U_3
\nonumber \\[0.2cm]
& + & S \big[ ih_0 - \sum_{i=1}^3 a_i U_i
+i \bar a_1 U_2 U_3 + i \bar a_2 U_1 U_3 + i \bar a_3 U_1 U_2 - \bar h_0 U_1 U_2 U_3 \big]
  \\[0.2cm]
& +  & \sum_{i=1}^3 T_i \big[ -ih_i  - \sum_{j=1}^3
U_j b_{ji}  + i U_2 U_3 \bar b_{1i}  + i U_1 U_3 \bar b_{2i} 
+ i U_1 U_2  \bar b_{3i} +  U_1 U_2 U_3 \bar h_i \big]   \ .
\nonumber
\label{wbtdual}
\eeqa
where the coefficients $h_i$, $b_{ij}$, ${\bar h}_i$, $\bar b_{ij}$, $i,j=1,2,3$ are the non-geometric flux quanta, and are related to the notation of \cite{Shelton:2005cf} in table 2 of \cite{Aldazabal:2006up}. These integers, and those controlling the 3-form fluxes, are subject to certain mutual consistency conditions, for whose discussion we refer to the references. 

We would however like to discuss an important compatibility conditions between the D-brane instantons of interest and the fluxes. Namely, the fluxes imply a gauging of certain isometries in the 4d effective theory, which must therefore not be violated by the non-perturbative effects. This is a generalization of the analysis in \cite{KashaniPoor:2005si}, and directly related to the Freed-Witten consistency conditions on brane wrappings in the presence of fluxes. In fact this issue had not appeared in our previous examples with 3-form fluxes because the pullback of the RR 3-form fluxes on the D3-brane 4-cycle vanished, hence satisfying the Freed-Witten consistency conditions (in fact this is generic for 4-cycles and 3-form fluxes on CY compactifications). In our present situation, the cancellation is not automatic, and if we fix the instanton D3-brane of interest, it restricts the set of fluxes compatible with it (or viceversa). These compatibility conditions are also identical to the Freed-Witten consistency conditions for 4d spacefilling D-branes wrapped on the same cycle as the instanton \cite{Camara:2005dc,Villadoro:2006ia}. In any event, they amount to requiring the invariance of the flux superpotential under the isometry along the modulus controlling the BPS phase of the relevant wrapped D-brane (for spacefilling branes, this is the gauging of the worldvolume $U(1)$ gauge symmetry). In our case we require invariance under
\beqa
T_1\to T_1+m_1 \lambda\quad ; \quad T_2 \to T_2+m_2\lambda
\eeqa
This invariance should hold for any value of the complex structure moduli $U_i$, and we obtain a set of constraints
\beqa
\sum_{i=1,2} m_i h_i\, =\, 0 \quad ; \quad 
\sum_{i=1,2} m_i b_{ji}\, =\, 0 \quad ; \quad 
\sum_{i=1,2} m_i \bar b_{ji}\, =\, 0 \quad ; \quad 
\sum_{i=1,2} m_i \bar h_i\, =\, 0 
\eeqa
These can be regarded as restrictions on the allowed D-brane instantons in a given flux compactification, or as restrictions on the allowed fluxes with one can turn on without destroying the existence of a given instanton. We adopt the latter viewpoint in our analysis. Clearly the above constraints still allow many flux configurations to be turned on, in a way compatible with our instantons of interest. Our main motivation for the above discussion, is to emphasize that the analysis of the compatibility of fluxes and instantons is amenable (as already mentioned in the references on the topic) to an analysis purely in terms of the 4d effective field theory, and fits perfectly with the viewpoint of the present paper.

It is clear that the superpotential (\ref{wbtdual}) contains interaction terms which are capable of contracting the diverse external legs in the original non-perturbative interaction (\ref{magneticfterm}). The result has a complicated dependence of the bulk and worldvolume flux quanta, although can be systematically computed for any given flux choice, and is in general nonzero. This corresponds in microscopic terms to the lifting of the $\ov\eta$ and $\ov\tau$ fermion zero modes by the flux background.
It would be interesting to explore the relation to spacetime susy breaking in this setup, which is expected from the general arguments in \cite{GarciaEtxebarria:2008pi}.
Leaving these aspects for future work, we close this section by hoping to have illustrated the use of the 4d effective action to discuss fermion zero mode lifting, even in situations with no available microscopic description.

\section{Worldsheet bulk-boundary map: \\ D-brane zero modes as zero modes of 4d moduli}
\label{sewing}

As is already clear from section \ref{exdef}, the matching of the microscopic fermion zero mode lifting and the effective field theory process of integrating out moduli, 
requires a very specific condition. Denote $G$ the flux component that lifts a given fermion zero mode, namely leads to a fermion zero mode coupling $G{\ov \eta}{\ov \eta}$ on the instanton action. Consider now the closed string moduli coupling to this fermion zero mode via  ${\ov\eta}{\ov{DU}}$. Then consistency with the effective field theory picture requires that the flux superpotential must contain a term $GUU$.

In other words, there must be a correlation between the couplings of the flux background to fermion zero modes and to spacetime moduli fields. For gauge field theory instantons, as in the example in section \ref{fieldtheory}, the relation between the couplings of instanton fermion zero modes and of spacetime fields is automatic, since the instanton fermion zero modes are particular components of the spacetime fields, localized on the core of the instanton. On the other hand, for general D-brane instantons in string theory, the instanton zero modes are open string fields, whose couplings are computed via disk diagrams, while the spacetime moduli fields are closed strings, whose couplings are computed via sphere diagrams. 
Still, this matching between specific open and closed modes, and their disk and sphere couplings, is found in particular examples, as in sections \ref{microview}, \ref{eftview}.

In the following we suggest an explanation for this matching, based on very general principles (as expected, since the existence of a matching is required on grounds of very general principles of effective field theory). Notice that the relevant couplings, of the flux to the fermion zero modes, of the fermion zero modes to the moduli, and of the flux to the moduli, enjoy powerful holomorphy properties, and indeed are computable in the corresponding topological string theory (similar to the discussion of other D-brane instanton couplings in \cite{Kachru:2008wt}). Therefore the appropriate setup for our question is open-closed topological string theory. In fact, the formulation of abstract properties of two-dimensional open-closed topological field theory \cite{Lazaroiu:2000rk} leads to an interesting correspondence between open and closed states and their correlators, the so-called bulk-boundary map, which we now explain.

Consider a topological string theory with open and closed strings. The closed string sector contains a set of states or vertex operators (which by abuse of language we do not distinguish), living on a Hilbert space $H_{\rm cl.}$. Let us consider a basis for this, with elements $\phi_i$, with respect to which one can expand any other state in $H_{\rm cl}$. On this space one can define the product $C(\phi,\phi')$
of two states $\phi$, $\phi'$ by scattering them into a third via a sphere diagram. In terms of the basis
\beqa
\phi_i\cdot \phi_j\, =\, C(\phi_i,\phi_j)\, =\,C_{ij}^{\, k} \, \phi_k
\eeqa
The closed string product coefficients $C_{ij}^{\, k}$ correspond to the sphere scattering amplitude of states $i$, $j$ into state $k$. 

One can operate similarly in the open string sector. For simplicity, we focus on the case of a single kind of boundary, i.e. a single kind of D-brane. One can introduce a basis in the Hilbert space $H_{\rm op.}$ of open strings, with elements $\psi_a$. One can use disk diagrams with two incoming states $\psi$, $\psi'$ scattering into an outgoing one to define an open string product $B(\psi,\psi')$. In terms of the basis we have
\beqa
\psi_a\cdot \psi_b\, =\, B(\psi_a,\psi_b)\, =\,B_{ab}^{\, c}\, \psi_c
\eeqa
Finally, there is a basic operation which defines a map (which is {\em not} an isomorphism) from the $H_{\rm cl.}$ to $H_{\rm op.}$. It corresponds to taking a disk diagram with one insertion of a closed string vertex operator $\phi$, and moving it to the boundary to define a boundary operator $e(\phi)$. Using the basis we have
\beqa
e(\phi_i)\, =\, e_i^{\, a}\, \psi_a
\eeqa
which defines the bulk-boundary map coefficients.
This process can be transformed (by a conformal transformation in the physical theory, or by deformation in the topological) into a diagram, denoted closed-open conduit, in which an incoming closed string state turns into an open one.

The sewing constraints in the open-closed topological field theory imply the bulk-boundary crossing symmetry formula \cite{Lazaroiu:2000rk}
\beqa
B(e(\phi),e(\phi'))\, =\, e(C(\phi,\phi'))
\label{crossing}
\eeqa
or in terms of the basis
\beqa
e_i^{\, a}\, e_j^{\, b}\, B_{ab}^{\, c}\, =\, e_{k}^{c}\, C_{ij}^{\, k}
\eeqa
The above is precisely the relation we hoped for! 

Let us begin explaining the right hand side. In our setup, the closed string fields $\phi$, $\phi'$ correspond to the spacetime moduli fields, for concreteness the modulinos \footnote{The structure of products and maps respects the $\IZ_2$ grading which allows to distinguish bosons and fermions, with their familiar rules in amplitudes.}. Their product $C(\phi,\phi')$ is the vertex operator of the flux component coupling to them. Its boundary image $e(C(\phi,\phi'))$ corresponds to the bulk flux restricted to the D-brane instanton. 

Consider now the left hand side. The operator $\psi=e(\phi)$ (and similarly for $\psi'=e(\phi')$) corresponds to an open string mode which couples, via the closed-open conduit, with the modulino $\phi$. Hence it corresponds to a D-brane instanton fermion zero mode, coupling to the modulino field. The product $B(\psi,\psi')$ corresponds to the field which couples to the fermion zero modes and lifts them. Equation (\ref{crossing}) implies that the object that lifts the worldvolume fermion zero modes is precisely the restriction to the D-brane instanton of the flux component coupling to the moduli fields.

Hence even for D-brane instantons with no gauge field theory, the bulk-boundary map allows to understand fermion zero modes as natural restrictions to the D-brane instanton of fields in the bulk. This works very simply in practice: consider the modulino fields in section \ref{eftview}, roughly speaking the structure in (\ref{modulinos}). Once restricted to the boundary, one should no longer distinguish between left and right. It is then possible to show that the structure of the modulino vertex operator becomes essentially that of the D-brane fermion zero modes, e.g. (\ref{fzms}) at the level of weight vectors.

The above discussion also sheds some light on the general question of the precise nature of D-brane instantons which are not gauge field theory instantons. Since their zero modes can be naturally regarded as excitations of closed string fields, they should be though of as instanton configurations of the closed string field background.

\section{Comments on the role of $\NN=1$ special geometry for D3-brane instantons}
\label{mixedhodge}

In this Section we make some general remarks on the relation of $\NN=1$ special geometry and mixed Hodge variations and the study of euclidean D3-brane instantons. 
In the literature \cite{Lerche:2002ck,Lerche:2002yw,Lerche:2003hs}, see also \cite{Jockers:2004yj,Jockers:2005zy}, these tools have been developed as a generalization of the special geometry in the complex structure moduli space of $\NN=2$ CY compactifications \cite{Strominger:1990pd}. The $\NN=1$ special geometry describes the moduli space of complex structure deformations and worldvolume deformations of B-type D-branes, which factorize only locally. The basic ingredients stemming from this special geometry are the coefficients of the open-closed chiral ring (closely related to the product coefficients in section \ref{sewing}), which are related to diverse 3-point amplitudes in disks and spheres. Given the close relation between certain disk diagrams for euclidean D3-branes on 4-cycles and 4d spacetime filling D7-branes on the same 4-cycles, it would be interesting to develop the tools of mixed Hodge variations to study aspects of D3-brane instanton effects.
We merely point out some interesting connections, emphasizing the differences with respect to the application to D7-branes. A systematic discussion is beyond the scope of the present work. Note that the language of $\NN=1$ special geometry is in principle tailored to allow the inclusion of 3-form flux superpotentials, thus presumably provides a framework to develop the more formal properties of the proposal in this paper.

Consider a type IIB compactification on a CY $X$ with D-branes wrapped on a holomorphic 4-cycle $S$. For simplicity, we focus the discussion in the case without orientifold planes, although the construction extends to orientifolded theories.
We are interested in studying the space of complex structure deformations of $X$ and the space of deformations of $S$ (keeping it holomorphic). At this point, it is important to distinguish if we are considering 4d spacefilling D7-branes or euclidean D3-brane instantons. In the case of D7-branes, the moduli of $S$ are 4d fields, and combine with the complex structure moduli of $S$ to form a joint moduli space ${\cal M}_{\NN=1}$, which does not factorize. In the case of euclidean D3-branes, the moduli of $S$ are bosonic zero modes of the instanton, over which one should integrate. The total space ${\cal M}_{\NN=1}$ described above, should not be regarded as a joint moduli space, but it remains an useful object, denoted ${\cal M}$ for short: it contain the information on how the instanton moduli space evolves as one moves around the complex structure moduli space ${\cal M}_{\NN=2}$ of $X$. This is possible because the space ${\cal M}_{\NN=1}$ is fibered over ${\cal M}_{\NN=2}$. The computation of the instanton amplitude at a given point of the base (for a given complex structure of $X$) implies the operation of integrating over the fiber (the moduli space of $S$). The $\NN=1$ special geometry of ${\cal M}$ presumably guarantees that the fibration structure is such that the result of the integration depends holomorphically on the base. 

This issue is thus related to the question of holomorphy of the non-perturbative superpotentials, as functions of the complex structure moduli (in the IIB picture). It thus continues a line of discussions in the papers  \cite{GarciaEtxebarria:2007zv,GarciaEtxebarria:2008pi,Gaiotto:2008cd}. As expected, since complex structure moduli space has a better behaved holomorphy properties in their coupling to D-branes (they couple holomorphically as coefficients in the D-brane superpotentials), as compared with Kahler moduli (which couple non-holomorphically, as Fayet-Illiopoulos terms, thus leading to real codimension one lines of marginal stability), the microscopic explanation of holomorphy of the superpotential does not require dramatic effects like multi-instantons, but rather a careful treatment of the instanton contribution as a function over moduli space, controlled in this case by ${\NN=1}$ special geometry.

A second aspect in which ${\NN=1}$ special geometry plays an important role is in determining the couplings of D3-brane instanton fermion zero modes. As discussed in \cite{Lerche:2003hs,Lerche:2002yw,Lerche:2003hs}, the spacetime superpotential on the moduli space ${\cal M}_{\NN=1}$ of complex structure moduli and D7-brane moduli, including flux superpotential and D7-brane moduli superpotential, is nicely encoded in the (relative) periods of a top holomorphic (relative) form, which depends on both closed and open string moduli. In suitable (flat) coordinates, the derivatives of the top relative form with respect to the different moduli provides the basic structure constants of the open-closed chiral ring (essentially the open-closed product coefficients). Therefore the $\NN=1$ special geometry contains the basic information about the couplings between D7-brane moduli and complex structure moduli. This information can be presumably translated to information on the couplings of D3-brane instanton fermion zero modes with complex structure moduli, of the kind we have used in our present work. We hope to come back to these interesting questions in future work.

\section{Conclusions}
\label{conclusion}

In this paper we have provided a description of the effects of fluxes on D-brane instantons, in terms of the 4d effective field theory of the compactification. This allows a description of this effect globally in moduli space, and a better understanding of some of its properties. The advantages of the effective field theory approach have been extensively discussed in the introduction and we will not repeat them here.

Although we have emphasized a pragmatic point of view, with concrete examples, the construction has interesting connection with more formal tools. For instance, the crucial requirement of the correlation between the couplings of instanton fermion zero modes and of 4d moduli fields is deeply rooted in the bulk-boundary map of topological string theory. This allows to make precise the intuition that instanton fermion zero modes are zero modes of 4d fields, localized at the core of the instanton. In fact, it suggests a very general correspondence between physics of instanton fermion zero modes, and their spacetime description, in general compactifications, with fluxes or without them. 

Finally, for the case of D3-brane instantons wrapped on 4-cycles, and their interplay with complex structure moduli of the Calabi-Yau compactification, there seem to be interesting connections with the ${\cal NN}=1$ special geometry and mixed Hodge structure in the joint moduli space ${\cal M}$. This is an interesting direction for further exploration. 

We hope our work provides an interesting step in the phenomenological and formal properties of combining fluxes and non-perturbative effects. 
\\
\begin{center}
{\bf Acknowledgements}
\end{center}
We thank I. Garcia-Etxebarria, L. Ib\'a\~nez, F. Marchesano, B. Schellekens for useful discussions. A.M.U. thanks M. Gonz\'alez  for encouragement and support. This work  has been supported by the Europea Commission under RTN European Programs MRTN-CT-2004-503369, MRTN-CT-2004-005105, by the CICYT (Spain) and the Comunidad de Madrid under project HEPHACOS  P-ESP-00346.  

\appendix

\section{Review of higher F-terms}
\label{bw}

In this section we briefly review some useful properties of multi-fermion F-terms (higher F-terms henceforth), following \cite{Beasley:2004ys,Beasley:2005iu}. We will adapt the discussion to our needs, considering the higher F-terms to be generated by BPS instantons with additional fermion zero modes. 

BPS instantons with $2p$ additional fermion zero modes, beyond the two $\NN=1$ goldstinos, generate a multi-fermion F-term of the form
\beqa
\label{fterm}
\delta S \,&= &\, \int \! d^4 x \, d^2 \theta \; \omega_{\ov i_1 \cdots
\ov i_p \,\ov j_1 \cdots \ov j_p} \, (\Phi) \; \left(\ov D_{\dot{\alpha}_1} \mskip 2
mu\ov\Phi{}^{\ov i_1} \ov D^{\dot{\alpha}_1}\mskip 2 mu\ov\Phi{}^{\ov
j_1}\right) \cdots \left(\ov D_{\dot{\alpha}_p}\mskip 2 mu\ov\Phi{}^{\ov i_p} 
\ov D^{\dot{\alpha}_p}\mskip 2 mu\ov\Phi^{\ov j_p}\right)\,, \nonumber \\
& \equiv & \, \int \! d^4 x \, d^2 \theta \; \CO_\omega
\eeqa
where the field dependent tensor $\omega_{\ov i_1 \cdots \ov i_p
\,\ov j_1 \cdots \ov j_p}$ is antisymmetric in the $\ov i_k$ and also in the $\ov j_k$, and symmetric under their exchange. Formally it can be regarded as a section of $\ov\Omega^p_{\cal M} \otimes \ov\Omega^p_{\cal M}$.

The conditions that $\delta S$ is supersymmetric and a non-trivial
F-term implies that $\omega$ belongs to a non-trivial cohomology class
in moduli space, for a certain cohomology, to which we refere as Beasley-Witten cohomology. For our purposes it is sufficient to consider the condition that the operator is supersymmetric, which amount to the statement that $\omega$ is annihilated by $\ov{\partial}$, {\em when regarded as a section of} $\ov\Omega^p_{\cal M} \otimes \Lambda^p T{\cal M}$  \cite{Beasley:2004ys,Beasley:2005iu}, namely
\beqa
\omega_{\ov i_1 \cdots \ov i_p \,\ov j_1 \cdots \ov j_p} \, =\, \omega_{\ov i_1 \cdots \ov i_p}^{\; \,j_1 \cdots  j_p} \, K_{j_1\ov j_1}\ldots K_{j_p\ov j_p}
\eeqa
where $K_{i\ov i}$ the Kahler metric in field space of chiral multiplets.

This condition is usually referred to as holomorphy of $\omega$. In some discussions in the literature, the lack of holomorphy of $\omega_{\ov i_1 \cdots \ov i_p \,\ov j_1 \cdots \ov j_p}$ due to the lowering of indices does not play an important role, and is considered only implicitly. 

Finally, let us mention that the behavior of higher F-terms from D-brane instantons across lines of marginal stability of the instantons, was discussed in \cite{GarciaEtxebarria:2008pi}. The higher F-terms remain continuous and holomorphic in a precise sense, and can discussed globally in moduli space, as we implicitly use in this paper.

\section{Sketch of the computation using supergraphs}
\label{supergraphs}

In this appendix we sketch the computation of the diagrams which provide the effective non-perturbative interactions once the relevant moduli (made massive by the flux superpotential) have been integrated out.

We consider the 4d theory with one chiral multiplet, a superpotential term and a 4-fermion F-term
\beqa
S\, =\, \int d^4x\, d^2\theta\, d^2{\ov\theta} \, K(U^i, {\ov U}^{\ov i})\, +\, 
\int d^4x\, d^2\theta \, \omega_{\ov i}^j \, K_{j\ov j}\, {\ov D}{\ov U}^{\ov i}\, {\ov D}{\ov U}^{\ov j}\, +\,   \int d^4x\, d^2\theta \, W(U)
\eeqa
Although the discussion is general, in our applications we are interested in situations where the higher F-term is of non-perturbative origin, arising from an instanton with additional fermion zero modes. The superpotential term will in our application be thought of as a flux superpotential, stabilizing the closed string modulus $U$. Along this line, we can consider the simple case where the superpotential is a mass term for the fields $U^i$. More in general, we can Taylor expand the superpotential and focus on the terms of the form $\partial_{ij} W U^i U^j$. As will be clear momentarily, higher orders will not modify the result we are after. 

\begin{figure}[!htp]
\centering
\psfrag{omegaK}{$\omega_{\ov i\ov j}$}
\psfrag{Ubari}{${\ov U}^{\ov i}$}
\psfrag{Ubarj}{${\ov U}^{\ov j}$}
\psfrag{Uk}{$U^k$}
\psfrag{Ul}{$U^l$}
\psfrag{partial}{$\partial_{kl} W$}
\includegraphics[scale=0.70]{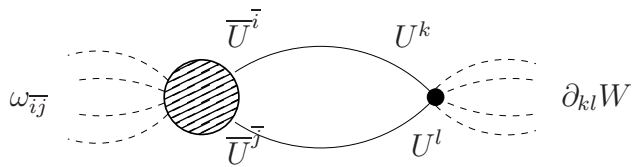}
\caption{\small The supergraph describing the integration out of the multiplets $U$.}
\label{fig:supergraph}
\end{figure}

We are interested in integrating out the massive fields $U^i$. To do so, we need to evaluate the Feynman diagrams where these fields run in internal lines. In particular, the diagram in Figure \ref{fig:supergraph} shows that the presence of the superpotential term can dress the higher F-term and produce a new non-perturbative superpotential contribution in the resulting low-energy effective field theory (in which $U$ has been integrated out). The computation can be made in the manifestly supersymmetric formalism of supergraphs. We refer to \cite{Grisaru:1979wc,Wess:1992cp} for the basic formalism, and simply recall some useful rules (for which we momentarily denote the chiral multiplets by $\Phi$ to keep the more familiar notation).

\begin{itemize}

\item Massless propagators $\langle \Phi^i {\ov \Phi}^{\ov j} \rangle$ are given by  $K^{i\ov j}\, \fund^{-1}$, where $K^{i\ov j}$ is the inverse Kahler metric in the space of chiral multiplet fields. Notice that we consider our perturbation theory as an expansion around a certain point in moduli space, and thus work in a kind of background field method, where the fields on which the propagator seems to depend are just replaced by the vevs around the point in moduli space.

\item  Since we deal with massless fields, the holomorphic $\langle \Phi\Phi \rangle$ or antiholomorphic ones $\langle {\ov \Phi} {\ov\Phi} \rangle$ vanish.

\item For each holomorphic field endpoint in an internal line one introduces a
factor of the square spinor derivative $-\frac 14 {\ov D}^2/4$ (and a factor of $-\frac 14 D^2$ for anti-holomorphic endpoint). This arises from the factors $-\frac 14{\ov D}^2$ in the functional derivatives with respect to the sources introduced to construct the generating functional
\beqa
\frac{\delta}{\delta J(x,\theta)}\, J(x',\theta')\, =\, -\frac 14 {\ov D}^2 \, \delta^4(x-x')\, \delta^4(\theta-\theta')
\eeqa

\item At each chiral vertex, one can absorb a factor of $-\frac 14{\ov D}^2$ to turn it into a full superspace interaction
\beqa
\int d^4x \, d^2\theta \, (-\frac 14 {\ov D}^2)\, f(\Phi)\, =\, \int d^4x\, d^2\theta\, d^2{\ov\theta} \, f (\Phi)
\eeqa

\end{itemize}

The supergraph of interest is shown in Figure \ref{fig:supergraph}, and can be computed as follows.
From the vertex associated to the higher F-term, we obtain two factors $-\frac 14 D^2$ from the ${\ov U}$ legs, and one of ${\ov D}^2$ from the vertex itself. The latter can
be used to promote the integration measure $d^2\theta\to d^4\theta$. In addition we have a factor $\omega_{\ov i}^{\ov m} K_{m\ov j}$, and a combinatorial factor of $1/2$. In total
\beqa
-\frac 18  \,\omega_{\ov i\ov j}\, D^2\, D^2 
\eeqa
Note that we commute things freely in the spirit of the background field method, namely we have functional dependences only on the vevs. From the vertex associated to the superpotential, we obtain two factors of $-\frac 14 {\ov D}^2$, one of which promotes the integral over half superspace to full superspace. In addition we have a factor $\partial_{kl}W$, and a combinatorial factor of $1/2$. Finally, the two propagators give factors $K^{\ov i k}/\fund$, $K^{\ov j l}/\fund$. The total result is
\beqa
-\frac 1{64} \frac{D^2{\ov D}^2D^2}{\fund\,{}^2} \, \omega_{\ov i\ov j} K_{\ov i k}K^{\ov j l} \, \partial_{kl}W =\,-\frac 14\, \omega_{\ov i\ov j} \, \partial^{\ov i\ov j} W \, \frac{D^2}{\fund}
\eeqa
where we have used $D^2{\ov D}^2D^2=16 \,\fund \,D^2$. We can further use
\beqa
\int d^4x\, d^4\theta \, (-\frac 14 D^2/\fund \,) \, f\, =\, \int d^4x\, d^2\theta \, f
\eeqa
and obtain a 4d effective interaction which we may write
\beqa
\int d^4x\, d^2\theta\,  K^{i{\ov i}}\, K^{j\ov j}\, \omega_{\ov i\ov j}\, \partial_{ij} W
\eeqa
This leads to the result mentioned in the main text, which agrees with that in \cite{Beasley:2004ys}, obtained using different arguments. Note that the superpotential in the above expression is not holomorphic throughout field space; however, as pointed out in \cite{Beasley:2004ys}, it need be holomorphic only when restricted to the moduli space of supersymmetric vacua. In our case, the only source of non-holomorphy involves the chiral multiplets which have been made massive, so the restriction to the massless sector indeed defines a holomorphic function.

The generalization of the argument to systems with more general higher F-terms is straightforward. Each microscopic superpotential term can be used to contract pairs of external legs in the higher F-term, reducing the number of external legs of the instanton interaction in pairs. It is easy to repeat the above supergraph arguments to obtain the final result explained in the main text. It is also easy to show that these are the only possible contraction patterns, and that there are no diagrams involving e.g. four internal lines between the higher F-term and a superpotential coupling. This is just the supersymmetric generalization of the fact that at each superpotential term there can be at most two fermionic legs.

\end{document}